\documentclass[envcountsame,runningheads,notitlepage]{llncs}

\usepackage{algorithm}
\usepackage{algpseudocode}
\usepackage{algorithmicx}
\usepackage{listings}
\usepackage{color}
\usepackage{amsmath}  
\usepackage{graphicx} 
\usepackage{amssymb} 
\usepackage{algpseudocode}
\usepackage{algorithm}
\usepackage{comment}
\usepackage{hyperref}
\usepackage{orcidlink}
\usepackage{pgfplots}
\usepackage[english]{babel}
\usepackage{float}
\floatstyle{plaintop}
\restylefloat{table}
\usepackage{pythonhighlight}
\usepackage{lscape}

\usepackage{graphicx}
\usepackage{epstopdf}
\usepackage{algorithm}
\usepackage{algpseudocode}
\AppendGraphicsExtensions{.gif}

\begin{document}

\title{Side Channel Analysis in Homomorphic Encryption}
\titlerunning{Side Channel Analysis in Homomorphic Encryption}

\author{
 Baraq Ghaleb\inst{1} \and William J Buchanan\inst{1} \orcidlink{0000-0003-0809-352}  }

\institute{Blockpass ID Lab, Edinburgh Napier University\\
  \href{mailto:b.buchanan@napier.ac.uk}{\{b.galeb, b.buchanan\}@napier.ac.uk} 
}  %

\maketitle

\begin{abstract}
Homomorphic encryption provides many opportunities for privacy-aware processing, including with methods related to machine learning. Many of our existing cryptographic methods have been shown in the past to be susceptible to side channel attacks. With these, the implementation of the cryptographic methods can reveal information about the private keys used, the result, or even the original plaintext. An example of this includes the processing of the RSA exponent using the Montgomery method, and where 0's and 1's differ in their processing time for modular exponentiation. With FHE, we typically use lattice methods, and which can have particular problems in their implementation in relation to side channel leakage. This paper aims to outline a range of weaknesses within FHE implementations as related to side channel analysis. It outlines a categorization for side-channel analysis, some case studies, and mitigation strategies.
\end{abstract}


\section{Introduction}
To enable secure processing of encrypted data, we require encryption schemes that support computation on ciphertexts. This can be achieved through Partial Homomorphic Encryption (PHE) or Fully Homomorphic Encryption (FHE). Encryption has been applied \emph{over-the-air} and \emph{at-rest}, but rarely during \emph{in-process} computations. To achieve this, we require encryption schemes that support computation on ciphertext. This is where Homomorphic Encryption (HE) comes into play, enabling computations on encrypted data without decryption. HE can be categorized into Partial Homomorphic Encryption (PHE) and Fully Homomorphic Encryption (FHE). PHE supports a limited set of arithmetic operations, while FHE enables arbitrary computations on encrypted data. The PHE methods include RSA, ElGamal, Paillier \cite{paillier1999public}, Exponential ElGamal, Elliptic Curve ElGamal \cite{elgamal1985public}, Paillier \cite{paillier1999public}, Damgard-Jurik \cite{damgaard2010generalization}, Okamoto–Uchiyama \cite{okamoto1998new}, Benaloh \cite{cohen1985robust}, Naccache–Stern \cite{naccache1997new}, and Goldwasser–Micali \cite{goldwasser2019probabilistic}. Overall, we can use RSA and ElGamal for multiplicative homomorphic encryption;  Paillier, Exponential ElGamal, Elliptic Curve ElGamal, Damgard-Jurik, Okamoto–Uchiyama, Benaloh and Naccache–Stern for additive homomorphic encryption; and Goldwasser–Micali for XOR homomorphic encryption.  

FHE, on the other hand, is typically based on lattice-based cryptography with modern implementations leverage libraries such as such as SEAL \cite{sealmanual} and OpenFHE \cite{openfhe_github}. While FHE introduces innovative approaches to privacy-preserving computation, its practical implementation remains relatively new. Although the underlying cryptographic methods are theoretically secure, real-world deployments often face challenges, particularly in mitigating side-channel vulnerabilities. These vulnerabilities can unintentionally leak information about encryption keys or even the original plaintext data, posing significant security risks.  This paper presents a comprehensive review of HE side-channel threats, highlighting key attack vectors and analyzing real-world case studies to assess their impact and potential countermeasures.. 

\subsection{Side-channel analysis }
\begin{figure}
    \centering
    \includegraphics[width=0.8\textwidth]{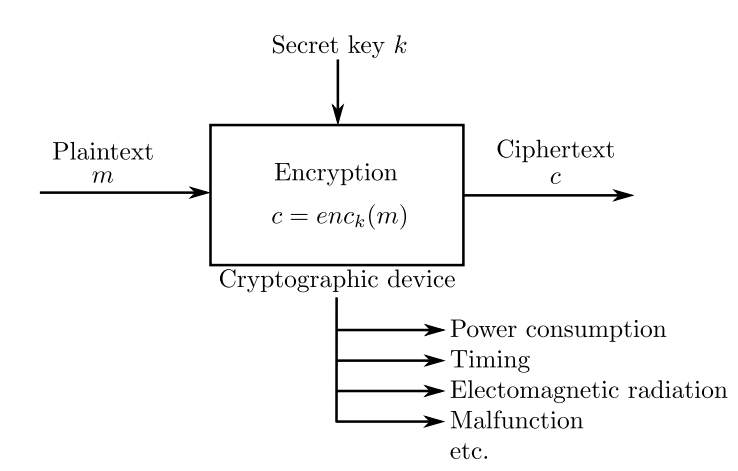}
    \caption{Possible side channels of a cryptographic device during an encryption \cite{strobel2009side}.}
    \label{fig:sidechannel}
\end{figure}

\begin{figure}[ht]
    \centering
    \includegraphics[width=0.8\textwidth]{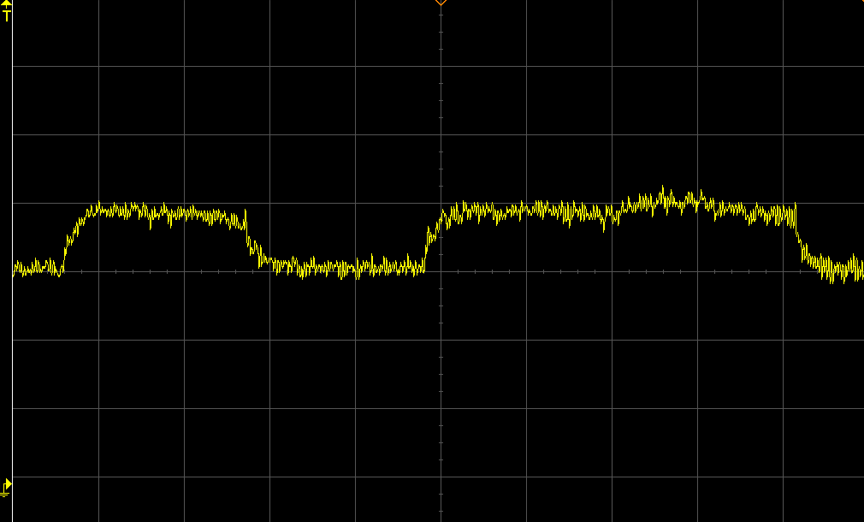}
    \caption{Observing RSA key bits using power analysis: The left peak shows the power consumption during the squaring-only step, the right (broader) peak shows the multiplication step, allowing exponent bits 0 and 1 to be distinguished \cite{power_image}.}
    \label{fig:power_analysis}
\end{figure}

Side-channel analysis (SCA) refers to a category of attacks that exploit information leaked from the physical implementation of cryptographic algorithms, rather than their  mathematical weaknesses  \cite{Mangard2010} \cite{Picek2023}. Unlike traditional cryptanalysis, SCA leverages observable characteristics of a system during operation such as timing variations, power consumption, or electromagnetic emissions \cite{Devi2021} \cite{strobel2009side} as depicted in Figure~\ref{fig:sidechannel}. These leaks can provide insights into sensitive data enabling attackers to bypass algorithmic defenses. For example, during the encryption process in the RSA algorithm, the ciphertext is generated by:
\begin{equation}
    c \equiv m^e \pmod{n}
\end{equation}
where \( c \), \( m \), \( e \), and \( n \) represent ciphertext, plaintext, key, and the product of \( p \) and \( q \), respectively. In practical implementations, modular exponentiation is commonly performed using the "Square and Multiply" algorithm, as shown in Algorithm~\ref{alg:square_multiply}. This method scans the exponent \( e \) bit by bit from left to right, performing an additional modular multiplication whenever a bit is set to 1. This variation causes detectable timing and power fluctuations, making it vulnerable to side-channel attacks where careful measurements and analysis can recover the secret key, bit by bit, as depicted in  Figure~\ref{fig:power_analysis}.

\begin{algorithm}
\caption{Exponentiation by Square-and-Multiply}\label{alg:square_multiply}
\begin{algorithmic}[1]
\State \( x \gets 1 \)
\For{each bit of \( e \) from left to right}
    \State \( x \gets x^2 \pmod{n} \)
    \If{current bit of \( e \) is 1}
        \State \( x \gets x \cdot m \pmod{n} \)
    \EndIf
\EndFor
\State \textbf{return} \( x \)
\end{algorithmic}
\end{algorithm}

 Homomorphic encryption (HE) faces unique security challenges due to side-channel vulnerabilities which arise for several reasons including:
\begin{itemize}
    \item \textbf{Computational Intensity:}  HE operations are highly computationally intensive, increasing exposure to side-channel analysis due to longer processing time \cite{Aydin2022}. 
    \item \textbf{Complex Operations:} HE schemes often involve complex mathematical operations with varying execution patterns that depending on the encrypted data or key material. These variations can result in distinguishable side-channel signatures \cite{Aydin2022}.
    \item \textbf{Cloud Computing Risk:} HE is frequently deployed in cloud environments, introducing additional attack vectors as malicious co-located processes may exploit leakage through shared hardware resources \cite{Onishi2024}.
\end{itemize}

\section{Background}
 In 1978, Rivest, Adleman, and Dertouzos \cite{rivest1978data}  were the first to introduce the concept of homomorphic operations, demonstrating its feasibility using the RSA cryptosystem. This early method supported multiply and divide operations \cite{asecuritysite_17070}, but lacked the ability to perform addition and subtraction.  Over time, HE  encryption evolved into two main categories: Partially Homomorphic Encryption (PHE), which supports a limited set of arithmetic operations, and Fully Homomorphic Encryption (FHE), which enables unrestricted addition, subtraction, multiplication, and division. Since Gentry introduced the first FHE scheme in 2009 \cite{homenc}, homomorphic encryption has evolved through four main generations:

\begin{itemize}
    \item 1st generation: Gentry’s method uses integers and lattices \cite{van2010fully} including the DGHV method.
    \item 2nd generation. Brakerski, Gentry and Vaikuntanathan’s (BGV) and Brakerski/ Fan-Vercauteren (BFV) leveraging the Ring Learning With Errors (RLWE) aproach \cite{brakerski2014efficient}.  These methods are similar to each other, and there is only a minor difference between them. They are generally used in applications with small integer values.
    \item 3rd generation: These include DM (also known as FHEW) and CGGI (also known as TFHE) and support the integration of  Boolean circuits for small integers. 
    \item 4th generation: CKKS (Cheon, Kim, Kim, Song) designed for efficient computations on floating-point numbers and can be applied to machine learning applications  as it can implement logistic regression methods and other statistical computation \cite{cheon2017homomorphic}.
\end{itemize}

\subsection{Public key or symmetric key}
\begin{figure}
  \includegraphics[width=\linewidth]{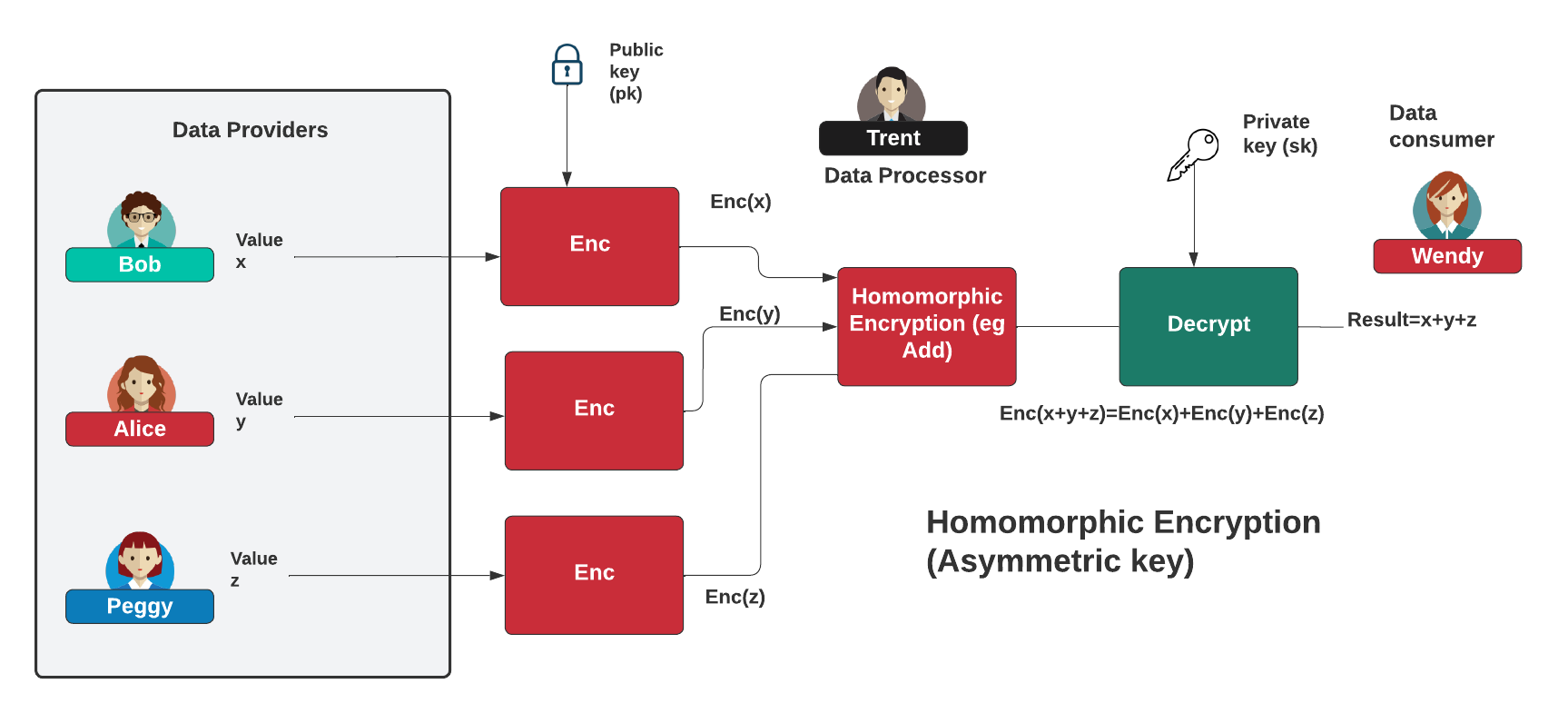}
  \caption{Asymmetric encryption (public key)}
  \label{fig:asym}
\end{figure}

\begin{figure}
  \includegraphics[width=\linewidth]{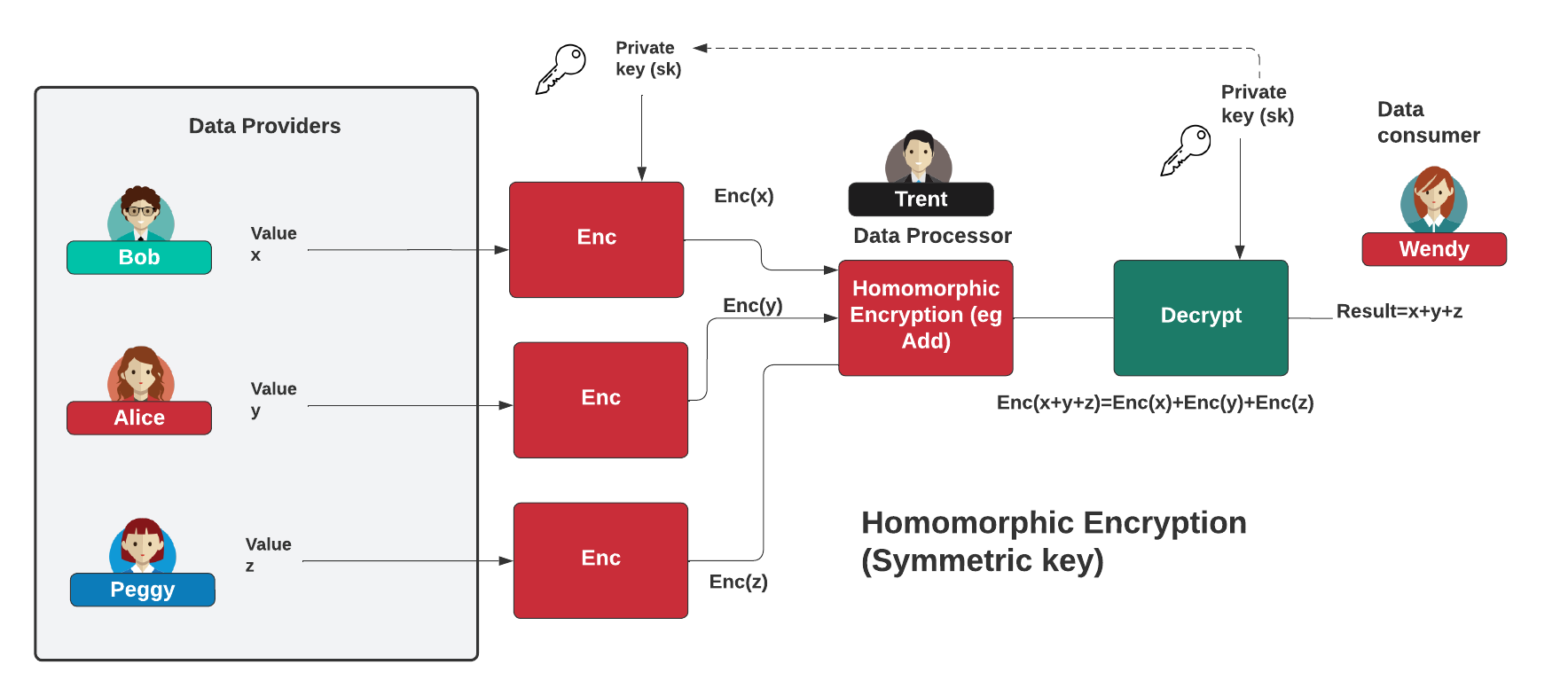}
  \caption{Symmetric encryption}
  \label{fig:sym}
\end{figure}

Homomorphic encryption can be implemented either with a symmetric key or an asymmetric (public) key. With symmetric key encryption, we use the same key to encrypt as we do to decrypt, whereas, with an asymmetric method, we use a public key to encrypt and a private key to decrypt.  In Figure \ref{fig:asym} we use asymmetric encryption with a public key ($pk$) and a private key ($sk$). With this Bob, Alice and Peggy will encrypt their data using the public key to produce ciphertext, and then we can operate on the ciphertext using arithmetic operations. The result can then be revealed by decrypting with the associated private key. In Figure \ref{fig:sym} we use symmetric key encryption, and where the data is encrypted with a secret key, and which is then used to decrypt the data. In this case, the data processor (Trent) should not have access to the secret key, as they could decrypt the data from the providers.

\subsection{Homomorphic libraries}

There are several homomorphic encryption (HE) libraries support FHE, including those optimized for CUDA and GPU acceleration. However, many are outdated or limited to a single encryption scheme. In practice, native language libraries are the most effective, as they enable direct compilation to machine code for better performance. The key languages for HE development are C++, Golang, and Rust, although some Python libraries exist through wrappers of C++ code such as HEAAN-Python, and its associated HEAAN library.  

One of the earliest libraries to support multiple homomorphic encryption schemes is \cite{asecuritysite_85691}, along with its variants SEAL-C\# and SEAL-Python. While it supports a wide range of methods, including BGV/BFV and CKKS, its development has slowed in recent years. It does, however, have support for Android and has a Node.js port \cite{asecuritysite_40933}. Another major library is OpenFHE, which was previously known as PALISADE. It is one of the most extensive FHE frameworks, offering support for multiple encryption schemes and advanced optimizations for real-world applications
For a more extensive list of homomorphic encryption libraries, refer to Wood et al. \cite{wood2020homomorphic}, which provides a comprehensive overview of available frameworks and their capabilities.
Within OpenFHE. The main implementations is this library are:

\begin{itemize}
    \item Brakerski/Fan-Vercauteren (\textbf{BFV}) scheme for integer arithmetic
    \item Brakerski-Gentry-Vaikuntanathan (\textbf{BGV}) scheme for integer arithmetic
    \item Cheon-Kim-Kim-Song (\textbf{CKKS}) scheme for real-number arithmetic (includes approximate bootstrapping)
    \item Ducas-Micciancio (\textbf{DM}) and Chillotti-Gama-Georgieva-Izabachene (\textbf{CGGI}) schemes for Boolean circuit evaluation.
\end{itemize}

\subsection{Bootstrapping}
\begin{figure}
  \includegraphics[width=\linewidth]{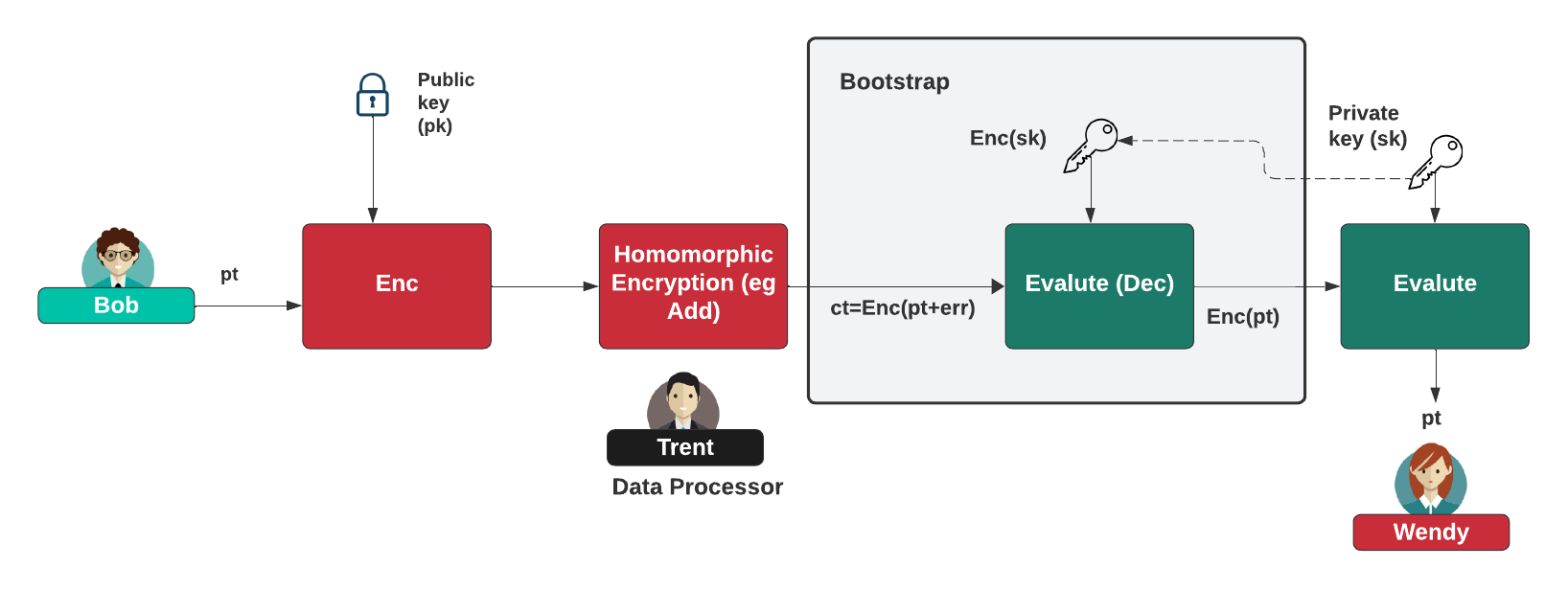}
  \caption{Bootstrapping}
  \label{fig:bootstrap}
\end{figure}

A fundamental concept in FHE is bootstrapping, which mitigates noise accumulation during computations. In Learning With Errors (LWE)-based schemes, noise is introduced to ensure security. Normally, encryption is performed with a public key, and decryption requires the corresponding private key. However, in bootstrapped FHE, an encrypted version of the private key is used to operate on the ciphertext, effectively reducing accumulated noise and restoring ciphertext usability. As illustrated in Figure \ref{fig:bootstrap}., bootstrapping involves evaluating decryption using an encrypted private key, after which the actual private key can be applied for final decryption.

The main bootstrapping methods are CKKS \cite{cheon2017homomorphic}, DM \cite{ducas2015fhew}/CGGI, and BGV/BFV. Overall, CKKS is generally the fastest bootstrapping method, while DM/CGGI is efficient with the evaluation of arbitrary functions. These functions approximate math functions as polynomials (such as with  Chebyshev approximation). BGV/BFV provides reasonable performance and is generally faster than DM/CGGI but slower than CKKS.

\subsection{Arbitrary smooth functions}
With approximation theory, it is possible to determine an approximate polynomial $p(x)$ that is an approximation to a function $f(x)$. A polynomial takes the form of $p(x)=a_n.x^n+a_{n-1}.x^{n-1}+...+a_1.x+a_0$, and where $a_0$... $a_n$ are the coefficients of the powers, and $n$ is the maximum power of the polynomial. In CKKS, arbitrary smooth functions can be efficiently approximated using the Chebyshev approximation \cite{al2023demystifying}, a method initially developed by Pafnuty Lvovich Chebyshev and nvolves the approximation of a smooth function using polynomials. Examples of these functions include $log_{10}$, $log_2$, $log_e$, and $e^x$  \cite{asecuritysite_65179}.

\subsection{Plaintext slots}
\begin{figure}
\centering
  \includegraphics[width=0.6\linewidth]{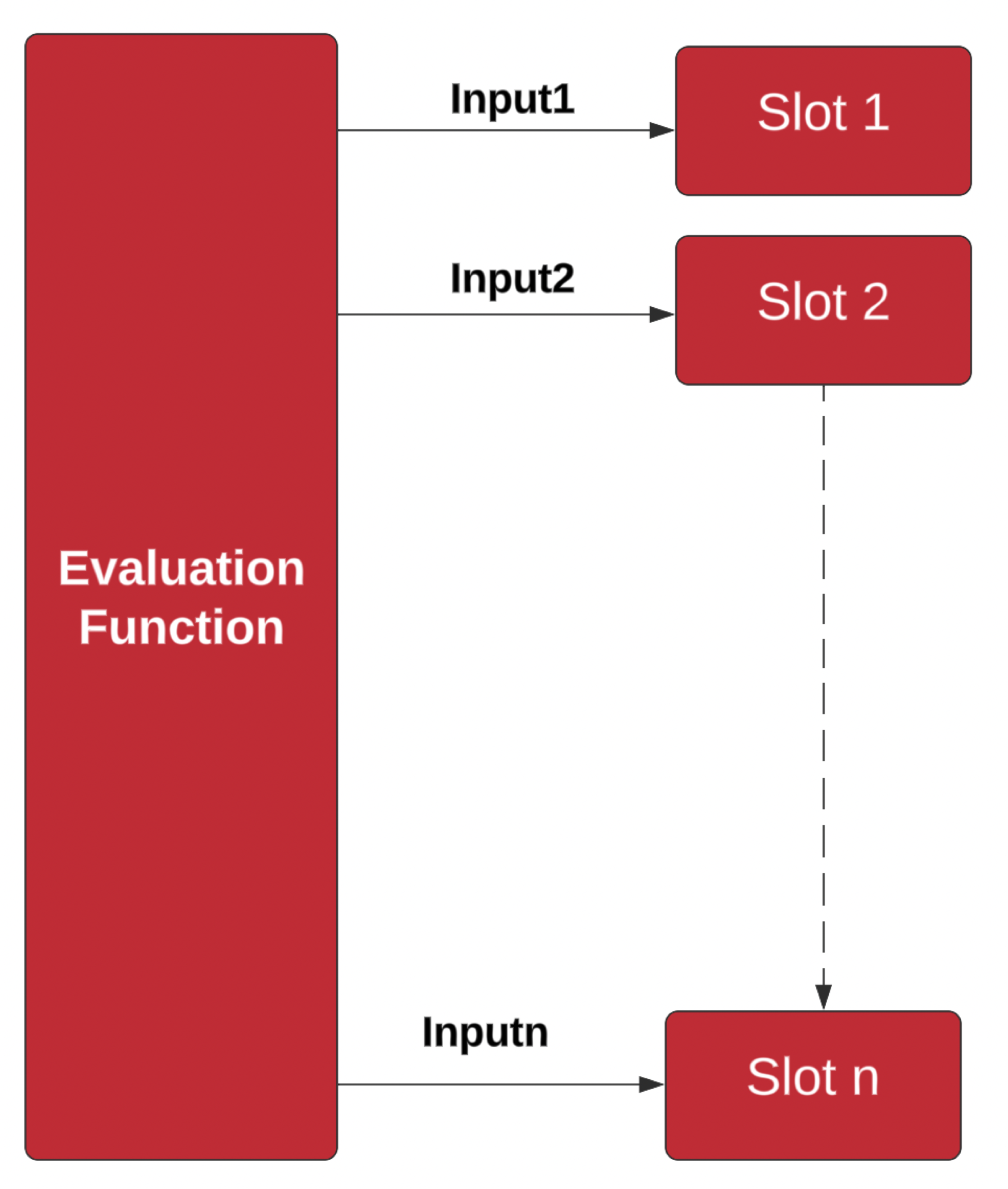}
  \caption{Slots for plaintext}
  \label{fig:slots}
\end{figure}

Many homomorphic encryption schemes support batch encryption, allowing multiple plaintext values to be packed into a single ciphertext. The number of values that can be encrypted together is referred to as the number of plaintext slots. This technique significantly improves computational efficiency by enabling parallel processing within a single encryption operation. Figure \ref{fig:slots} illustrates the concept of plaintext slots and their role in optimizing homomorphic computations.

\subsection{BGV and BFV}
Both BGV and BFV homomorphic encryption schemes utilize the Ring Learning With Errors (RLWE) method \cite{brakerski2014efficient}.  In BGV, we define a moduli ($q$), which constrains the range of the polynomial coefficients. These schemes employ a moduli hierarchy, where different levels of modulus are used to manage precision and computational efficiency. The encryption process starts by defining the finite group $\mathbb{Z}_q$,and constructing a polynomial ring  by dividing our operations with $(x^n+1)$ and where $n-1$ is the largest power of the coefficients. The message can then be represented in binary as:

\begin{equation}
m=a_{n-1}a_{n-2}...a_0
\end{equation}

This is converted into a polynomial form:

\begin{equation}
\mathbf{m}=a_{n-1} x^{n-1} + a_{n-2} x^{n-2}+...+a_1 x + a_0 \pmod q
\end{equation}

The polynomial’s coefficients form a vector representation of the plaintext. For improved efficiency, messages can also be encoded in ternary (such as with -1, 0 and 1). Finally, the plaintext modulus is defined as follows:

\begin{equation}
t = p^r
\end{equation}

and where $p$ is a prime number and $r$ is a positive number. We can then define a ciphertext modulus of $q$, and which should be much larger than $t$. To encrypt with the private key of $\mathbf{s}$, we implement:

\begin{equation}
(c_0, c_1) =\left( \frac{q}{t}.\mathbf{m}  + \mathbf{a}.\mathbf{s} + e,\mathbf{-a} \right) \mod q
\end{equation}

To decrypt:

\begin{equation}
m = \bigl \lfloor \frac{t}{q}(c_0+c_1).\mathbf{s} \bigr \rceil
\end{equation}

This works because:

\begin{align}
m_{recover} &= \bigl \lfloor  \frac{t}{q}\left(\frac{q}{t}.\mathbf{m}  + \mathbf{a}.\mathbf{s} + e -\mathbf{a}.\mathbf{s} \right) \bigr \rceil\\
&= \bigl \lfloor \left( \mathbf{m}  + \frac{t}{q}.e  \right) 
 \bigr \rceil\\
 & \approx m \\
\end{align}

For two message of $m_1$ and $m_2$, we will get:

\begin{align}
Enc(m_1+m_2) &= Enc(m_1) + Enc(m_2)\\
Enc(m_1.m_2) &= Enc(m_1) . Enc(m_2)
\end{align}

\subsubsection{Noise and computation}

Each addition or multiplication increases error, necessitating bootstrapping to reduce noise. While addition and plaintext/ciphertext multiplication are relatively fast, ciphertext/ciphertext multiplication is more computationally intensive and introduces the most noise. Bootstrapping remains the most demanding operation in the process.

\subsubsection{Parameters}
To balance security, precision, and efficiency, the ciphertext modulus $q$ and the plaintext modulus $t$ carefully chosen. Both of these are typically to the power of 2 with $t$ defining the plaintext space and $q$ determining the noise-handling capacity.  For example, $q$ of $2^{240}$ and $t$ of 65,537 provide sufficient precision while keeping computation feasible. As the value of $2^q$ is likely to be a large number, we typically define it as a $log\_q$ value.  Thus, a ciphertext modulus of $2^{240}$ will be 240 as defined as a $log_q$ value.

\subsubsection{Public key generation}
We select the private (secret) key using a random ternary polynomial (-1, 0, and 0 coefficients) which has the same degree as our ring. The public key is then a pair of polynomials as:

\begin{align}
\mathbf{pk_1} &= (\mathbf{r}.\mathbf{sk}+e) \pmod q\\
\mathbf{pk_2}  &= \mathbf{r}
\end{align}

Where $r$ is a random polynomial value. To encrypt with the public key ($pk$):

\begin{equation}
(c_0, c_1) =\left( \frac{q}{t}.\mathbf{m}  + \mathbf{a}.\mathbf{s}.\mathbf{r}  + e,\mathbf{-a}.\mathbf{r} \right) \mod q
\end{equation}

We then decrypt with the private key ($s$);

\begin{equation}
m = \bigl \lfloor \frac{t}{q}(c_0+c_1).\mathbf{s} \bigr \rceil
\end{equation}
This works because:

\begin{align}
m_{recovered} &= \bigl \lfloor  \frac{t}{q}\left(\frac{q}{t}.\mathbf{m}  + \mathbf{a}.\mathbf{s}.\mathbf{r} + e -\mathbf{a}.\mathbf{r}.\mathbf{s}. \right) \bigr \rceil\\
&= \bigl \lfloor \left( \mathbf{m}  + \frac{t}{q}.e  \right) 
 \bigr \rceil\\
 & \approx m \\
\end{align}

\subsection{The HEAAN library of CKKS}

HEAAN (HE for Arithmetic of Approximate Numbers) is a HE library based on the CKKS scheme. The CKKS enables approximate arithmetic over complex numbers \cite{cheon2017homomorphic}  and is a levelled approach that involves the evaluation of arbitrary circuits of bounded (pre-determined) depth. These circuits can include ADD (X-OR) and Multiply (AND).

HEAAN uses a rescaling procedure to manage plaintext size, applying approximate rounding by truncating the ciphertext to a smaller modulus. This approach enables efficient parallel encryption computations. However, as operations progress, the ciphertext modulus can shrink to a point where further computation becomes impossible.

The CKKS scheme, used in HEAAN, performs approximate arithmetic over complex numbers ($\mathbb{C}$). It introduces an encryption error that blends with computational noise inherent in approximate calculations. The encryption process converts a plaintext message (\textit{M}) to a cipher message (\textit{ct}) using a secret key \textit{sk}. To decrypt ([⟨\textit{ct},\textit{sk}⟩]\textit{q}), we produce an approximate value with a small error (\textit{e}).

\subsubsection{Chebyshev approximation}
With approximation theory, it is possible to determine an approximate polynomial $p(x)$ that is an approximation to a function $f(x)$. A polynomial takes the form of $p(x)=a_{n}.x^{n}+ a_{n-1}.x^{n-1}+ a_1.x + a_0$, and where $a_0 ... a_n$ are the coefficients of the powers, and $n$ is the maximum power of the polynomial. In this case, we will evaluate arbitrary smooth functions for CKKS and use Chebyshev approximation. These were initially created by Pafnuty Lvovich Chebyshev. This method involves the approximation of a smooth function using polynomials.

Overall, with polynomials, we convert our binary values into a polynomial, such as 101101 is:

\begin{equation}
x^5+x^3+x^2+1
\end{equation}

Our plaintext and ciphertext are then represented as polynomial values.

\subsubsection{Approximation theory}

With approximation theory, we aim to determine an approximate method for a function f(x). It was Pafnuty Lvovich Chebyshev who defined a method of finding a polynomial p(x) that is approximate for f(x). Overall, a polynomial takes the form  of:

\begin{equation}
p(x)=a_n.x^n+a_{n-1}.x^{n-1}+a_1.x+a_0
\end{equation}

and where $a_0 ... a_n$ are the coefficients of the powers, and $n$ is the maximum power of the polynomial. Chebyshev published his work in 1853 as "Theorie des mecanismes, connus sous le nom de parall´elogrammes". His problem statement was “to determine the deviations which one has to add to get an approximated value for a function $f$, given by its expansion in powers of $x-a$, if one wants to minimise the maximum of these errors between $x = a - h$ and $x = a + h$, $h$ being an arbitrarily small quantity".

\subsection{Polynomial evaluations}
A polynomial takes the form form of \(p(x)=a_{n}.x^{n}+ a_{n-1}.x^{n-1}+ a_1.x + a_0\), and where $a_0 ... a_n$ are the coefficients of the powers, and \(n\) is the maximum power of the polynomial. With CKKS in OpenFHE, we can evaluate the result of a polynomial for a given range of $x$ values. For example, if we have $p(x)=5.x^2+3.x+7$ will give a result of $p(2)= 33$.

\section{Categories of Side-Channel Analysis}
Side-channel analysis (SCA) is a powerful technique used by attackers to extract sensitive information from cryptographic implementations by observing physical characteristics like power consumption, timing information, electromagnetic radiation, or error responses. Eavesdroppers can monitor the power consumed during operations, the electromagnetic radiation emitted during decryption and signature generation, or the time taken to perform cryptographic operations. By analyzing these physical signals, attackers can infer secret information, such as cryptographic keys, and compromise the security of the system. Additionally, attackers can exploit how a cryptographic device behaves when errors occur, potentially revealing vulnerabilities in the implementation \cite{Devi2021} \cite{Spreitzer2018}. Side-channel attacks can be broadly categorized into two main types: \textbf{passive side-channel attacks} (also known as tamper attacks) and \textbf{active side-channel attacks}. Each type exploits physical information leakage in distinct ways \cite{Devi2021}:

\subsection{Passive Side-Channel Attacks}
Passive side-channel attacks focus on observing physical signals emitted by a device during its normal operation without interfering with the device itself. Passive attacks are further divided into two subcategories \cite{Standaert2010}:

\begin{itemize}
    \item \textbf{Simple Analysis Attacks:}  
    These attacks rely on direct observations of leaked signals, such as power consumption or timing information. The attacker interprets these signals to deduce sensitive data, often requiring minimal computational effort.Popular examples of this are Simple Power Analysis (SPA), Simple Electromagnetic Analysis (SEMA), and Basic timing analysis

    \item \textbf{Differential Analysis Attacks:}  
    These attacks involve statistical analysis of multiple measurements to extract subtle correlations between the physical characteristics and the secret data. Differential Power Analysis (DPA) is a well-known example of this type of attack, where the attacker correlates power traces with known plaintext or ciphertext to recover secret keys.
\end{itemize}

\subsection{Active Side-Channel Attacks}
Active side-channel attacks involve intentionally manipulating the target device to induce faults or abnormal behavior. By analyzing the device's response to such tampering, attackers can extract critical information, such as secret keys. These attacks exploit vulnerabilities in how cryptographic devices handle unexpected conditions. Side-Channel Attacks could also be  classified based on the type of information channel exploited as follows.

\subsection{Power Analysis}
Power analysis attacks exploit variations in power consumption during cryptographic operations and they could be categorized to the following key classes \cite{Messerges2000}:
\begin{itemize}
\item Simple Power Analysis (SPA):
Simple Power Analysis (SPA) is a technique that involves directly analyzing power consumption measurements captured during encryption and decryption operations. By visually inspecting these power traces, attackers can often identify patterns that reveal information about the cryptographic algorithm's execution flow or even the secret key itself. This technique was pioneered by Paul Kocher and his colleagues, Jaffe and Jun, who also introduced Differential Power Analysis (DPA) in their influential paper. SPA can yield information about a device's operation as well as key material. SPA could be used to collect information about the target's cryptographic implementations by, e.g., interpreting how many rounds are used during encryption/decryption. SPA is the simplest form of power analysis.

\item Differential Power Analysis (DPA):
Differential Power Analysis (DPA) is a sophisticated side-channel attack technique that involves statistically analyzing power consumption data collected during cryptographic operations. Unlike Simple Power Analysis (SPA), which relies on direct observation of power traces, DPA examines subtle correlations between power consumption and processed data over multiple cryptographic operations. In a DPA attack, the adversary uses hypothetical power models, such as the Hamming weight or Hamming distance, to predict power consumption patterns based on specific inputs or intermediate values. By comparing these predictions with actual power traces, the attacker can identify correlations and deduce sensitive information, such as cryptographic keys. DPA is particularly powerful because it can extract information even in the presence of noise and countermeasures, making it a significant threat to devices with inadequate side-channel protections.

\item Correlation Power Analysis (CPA):
Correlation Power Analysis (CPA) is an advanced side-channel attack technique that uses statistical correlation (e.g.  Pearson correlation coefficient) to link power consumption patterns of a cryptographic device with hypothetical models, such as Hamming weight or Hamming distance. By analyzing multiple power traces recorded during cryptographic operations and comparing them to predicted power values for various key guesses, CPA identifies the key hypothesis with the highest correlation as the correct one. Its robustness to noise and precision make CPA a significant threat to cryptographic devices, necessitating countermeasures like randomization, noise injection, or constant power hardware designs to mitigate its effectiveness.
\end{itemize}

\subsection{Timing Analysis}
Timing analysis attacks exploit variations in the execution time of cryptographic operations to extract sensitive information \cite{Devi2021}. By analyzing the relationship between timing fluctuations and the operations being performed, attackers can infer details about the cryptographic process. These attacks often target subtle discrepancies in hardware or software behavior. Common techniques include:
\begin{itemize}
\item Monitoring cache access patterns to infer data dependencies and memory usage.
\item Analyzing branch prediction behavior to deduce control flow or conditional operations.
\item Examining instruction scheduling delays to uncover computational bottlenecks or specific algorithmic steps.
\end{itemize}

\subsection{Electromagnetic Analysis}
Electromagnetic (EM) analysis leverages the electromagnetic emissions generated by a device during operation to extract or disrupt sensitive information \cite{Spreitzer2018}. This technique is versatile, with both passive observation and active interference methods.

\subsection{Fault Analysis}
Fault analysis, a sophisticated form of active attack, exploits errors intentionally introduced into cryptographic computations to extract sensitive information, such as secret keys or intermediate values \cite{Devi2021}. Attackers manipulate the system's environment or operational conditions to disrupt its normal behavior. By analyzing the resulting faulty outputs, they can infer critical data through techniques like differential fault analysis or other statistical methods. Common fault injection methods include \cite{Spreitzer2018}:
\begin{itemize}
\item Voltage Glitching: Intentionally manipulating the power supply voltage to create transient voltage spikes or dips, causing the circuit to malfunction and potentially reveal sensitive information.
\item Clock Manipulation: Altering the frequency or phase of the clock signal to disrupt the timing of operations, leading to incorrect calculations and potential security vulnerabilities.
\item Laser/EM Injection: Employing focused laser beams or electromagnetic radiation to target specific circuit components and induce faults, such as bit flips or temporary circuit failures.
\end{itemize}

\section {Case Studies: Side-channel Attacks on HE Systems}
 In the context of homomorphic encryption (HE), SCA poses unique challenges. While HE theoretically ensures robust cryptographic security, its practical implementations may leak exploitable side-channel information, particularly during computationally intensive operations like key generation, encryption, and homomorphic evaluation.

\subsection{Single-Trace Attack on SEAL’s BFV Encryption Scheme}

The study in \cite{Aydin2022} introduced the first single-trace side-channel attack on homomorphic encryption (HE), specifically targeting the SEAL implementation of the Brakerski/Fan-Vercauteren (BFV) scheme prior to v3.6. The attack exploits power-based side-channel leakage during the Gaussian sampling phase of SEAL’s encryption process, enabling plaintext recovery with a single power measurement. The proposed attack follows a four-step methodology to recover plaintext messages from SEAL’s BFV encryption scheme. First, the attack identifies each coefficient index being sampled during the encryption process. Second, it extracts sign values from control-flow variations, which reveal key information about the sampled coefficients. Third, the attack recovers the coefficients with high probability using data-flow variations observed in the power trace. Finally, the Blockwise Korkine-Zolotarev (BKZ) algorithm is applied to explore and estimate the remaining search space, further refining the attack's success.
The study also focused on recovering plaintext messages by extracting coefficients of error polynomials. These coefficients are integral to the encryption process, and their recovery undermines the cryptographic hardness of the scheme. Hence, the methodology identifies and exploits vulnerabilities in the \texttt{set\_poly\_coeffs\_normal} function, responsible for error polynomial sampling. These vulnerabilities include branch operations that reveal sign information, and negation operations that reduce false positives by exploiting Hamming weight differences. Using real power measurements on a RISC-V FPGA implementation of SEAL v3.2, the attack reduces the security level of the plaintext encryption from $2^{128}$ to $2^{4.4}$, highlighting the significant implications of side-channel attacks on HE.   `However, the attack has several drawbacks, including the need for profiling and key configuration, requiring many traces to create accurate templates \cite{Aydin2022}. It is also limited to a single device, and cross-device attacks may need machine learning for profiling. Furthermore, the attack was performed at a 1.5 MHz frequency, as higher frequencies increase noise and may require advanced equipment. Additionally, targeting more secure versions (196-bit or 256-bit keys) is harder due to increased precision and more coefficients.

\subsection{Single-Trace ML Attack on CKKS SEAL's Key Generation}
The authors in \cite{Aydin2024} uncover new side-channel vulnerabilities in Microsoft SEAL by focusing on its number theoretic transform (NTT) function using power analysis.  Specifically, it presents an attack targeting the NTT operation within the SEAL CKKS scheme’s key generation process. The study demonstrates that the NTT, used during key generation, leaks ternary values ($-1$, $0$, $+1$) corresponding to secret key coefficients. The key innovation lies in developing a sophisticated two-stage neural network-based classifier capable of extracting side-channel information from a single measurement, demonstrating an unprecedented 98.6\% accuracy in revealing secret key coefficients on the ARM Cortex-M4F processor. The research explores the impact of compiler optimizations, analyzing SEAL’s NTT implementation across optimization levels from \texttt{-O0} (no optimization) to \texttt{-O3} (maximum optimization). While \texttt{-O3} eliminates previously identified vulnerabilities, the study reveals new side-channel leakages in the \texttt{guard} and \texttt{mul\_root} operations under this setting. Random delay insertion, evaluated as a countermeasure, is shown to be ineffective against the proposed attack. This study distinguishes itself from prior side-channel analyses by targeting the latest version of SEAL (v4.1) and addressing vulnerabilities absent in other implementations. Unlike multi-trace attacks, which focus on decryption operations, this single-trace attack directly exploits NTT computations, bypassing defenses like masking. The study also reveals additional leakages with \texttt{-O3} optimization and refines attack accuracy. The findings underscore the need for robust countermeasures to secure FHE implementations like SEAL against single-trace side-channel attacks, particularly those targeting efficient and constant-time arithmetic.

\subsection{Side-Channel Vulnerabilities in LWE/LWR-Based Cryptography}
The authors in \cite{Ravi2022} demonstrated successful attacks against multiple post-quantum cryptography implementations, breaking through various side-channel countermeasures, including masking and shuffling techniques. Their work revealed that these vulnerabilities are not implementation-specific but rather stem from core algorithmic properties of LWE/LWR-based cryptography. While this research does not explicitly target homomorphic encryption schemes, its findings have significant implications for the broader family of LWE/LWR-based cryptographic systems, including homomorphic encryption. As many modern homomorphic encryption schemes are built upon these same mathematical foundations, they could potentially be vulnerable to similar side-channel attacks targeting specifically, the incremental storage of decrypted messages in memory and ciphertext malleability properties inherent to LWE/LWR-based schemes. 

\subsection{Cache-Timing Attack on the SEAL Homomorphic Encryption Library}
The authors in \cite{Cheng2022} exposed a cache-timing vulnerability in the SEAL homomorphic encryption library, specifically in its implementation of Barrett modular multiplication. The researchers identified a timing side-channel in the non-constant-time implementation of extra-reductions using the ternary operator, which leaks information about the secret key. Leveraging a novel remote cache-timing methodology, the attack aims to recover the secret key involved in modular multiplications. By analyzing ciphertexts that cause an extra reduction, the researchers solve Diophantine equations to progressively narrow down the range of possible secret keys. The approach combines insights from Bézout's theorem with optimized enumeration techniques to refine key candidates efficiently. The attack requires as few as eight ciphertexts that trigger extra reductions to uniquely determine the secret key. To eliminate the vulnerability, the authors proposed a constant-time implementation of the \texttt{SEAL\_COND\_SELECT} macro. This replacement avoids conditional branching by using bitwise operations to compute results in constant time, ensuring that the timing of the operation no longer depends on the input values.

\subsection{Single-Trace Side-Channel Attacks on Masked Lattice-Based Encryption}

The authors in \cite{Primas2017} introduced a single-trace side-channel attack targeting lattice-based cryptography, demonstrating its vulnerability to key recovery using observations from a single decryption. Unlike previous side-channel attacks, this approach is uniquely powerful because it can penetrate even masked implementations by recovering individual shares and subsequently reconstructing the complete decryption key. It focuses on the Number Theoretic Transform (NTT), a critical component in almost all efficient lattice-based cryptography implementations, making the attack applicable across a broad range of encryption schemes, including homomorphic encryption schemes. Unlike previous differential power analysis (DPA) attacks, which overlooked NTT, the authors leverages its less-protected nature. The attack comprises three main steps: (1) side-channel template matching on modular operations during the inverse NTT; (2) combining intermediate probabilities via belief propagation (BP) on the FFT-like NTT structure, optimized to make BP computationally feasible; and (3) utilizing leaked intermediate values along with the public key to recover the private key through lattice decoding. 

\subsection{A Practical Full Key Recovery Attack on TFHE and FHEW by Inducing Decryption Errors}

The study in \cite{Chaturvedi2022} introduced the latest side-channel attack targeting Fully Homomorphic Encryption (FHE) schemes on the server side.  Unlike prior attacks that focused on the client side, this study demonstrates that a malicious server can inject carefully calculated perturbations into ciphertexts stored on the cloud to induce decryption errors on the client side. The client, unaware of the malicious intent, reports these errors to the server. By analyzing the pattern of errors and the timing information associated with homomorphic operations, the attacker can gradually extract the underlying error values for each ciphertext. Specifically, these errors are used to reconstruct a system of linear equations that, when solved, compromise the security of the underlying Learning with Errors (LWE) problem and recover the client's secret key. By strategically inducing errors and using a binary-search approach to recover precise error values, the researchers successfully developed a technique to extract secret keys with a remarkably low number of client queries to avoid detection. By leveraging timing information during homomorphic gate computations, the authors significantly reduce the number of queries needed to extract errors, achieving efficient key recovery. The attack successfully recovered secret keys for two widely used FHE libraries, FHEW and TFHE, requiring 8 and 23 queries per ciphertext error extraction, respectively. The full-key recovery attack was demonstrated on practical scenarios with TFHE (key size: 630 bits) and FHEW (key size: 500 bits), involving 19,838 and 7,565 client queries, respectively.

\section {Mitigation Strategies for HE Side-channel Attacks}
As mentioned earlier,  HE is not immune to side-channel attacks, which exploit implementation-specific vulnerabilities such as timing information, power consumption, or electromagnetic emissions to infer sensitive data. Addressing these threats is critical to ensuring the practicality and trustworthiness of HE systems. This section explores various mitigation strategies designed to harden HE implementations against side-channel attacks and to  reduce the attack surface, bolstering the overall security of HE-based applications.

\subsection{Constant-Time Implementations}
Constant-time implementations are a cornerstone of secure cryptographic design, addressing timing side-channel vulnerabilities by ensuring that operations execute in a consistent manner, independent of the input data or secret parameters \cite{Lyu2018}. Cryptographic algorithms and their implementations are crafted to eliminate timing variations that could inadvertently leak sensitive information, such as secret keys or computational states. This involves uniform execution patterns, where each branch of the code consumes the same computational resources and takes an equal amount of time to complete, regardless of the processed data. By adhering to constant-time principles, developers can significantly reduce the risk of timing attacks, where adversaries analyze variations in execution time to extract critical cryptographic secrets.
However, achieving true constant-time behavior can be challenging in complex cryptographic systems, as even minor variations in hardware architectures or compiler optimizations may introduce unintended inconsistencies. Constant-time implementations can vary across platforms. For instance, the "constant-time" fix in OpenSSL designed to mitigate the Lucky Thirteen attack still shows data-dependent execution times on ARM architectures \cite{Lyu2018} \cite{Cock2014}. Additionally, constant-time designs often come with a performance overhead, as they may require additional computations or stricter coding practices to ensure uniformity. This can impact the efficiency of Homomorphic Encryption systems, where computational performance is already a critical concern.

\subsection{Masking and Blinding Techniques}

Masking and blinding are techniques that randomize intermediate values to mitigate side-channel attacks. When applied to symmetric block ciphers, this is referred to as masking, where data is split into randomized shares. In contrast, when used in public key cryptosystem implementations, it is called blinding, involving the addition of random noise to obfuscate correlations \cite{Hou2024}.  These techniques introduce random noise or perturbations during cryptographic operations to obscure the computational process and prevent attackers from correlating power traces, timing variations, or other measurable characteristics with secret data. By randomizing internal computations and intermediate values, they aim to reduce the leakage of sensitive information during cryptographic transformations. However, such defenses have significant limitations, particularly their susceptibility to single-trace side-channel attacks \cite{Aydin2022} \cite{Cheng2022}. In such scenarios, attackers can exploit advanced statistical or machine learning techniques to bypass the randomness introduced by masking and extract secret information from a single observation. As a result, while masking/blinding can provide a basic level of protection, it is not a recommended standalone defense and should be complemented by more robust strategies to ensure comprehensive security against side-channel threats.

\subsection{Shuffling and Randomization Techniques}

Shuffling and randomization \cite{Patranabis2016Shuffling}  techniques can play a crucial role in mitigating side-channel attacks in the context of Homomorphic Encryption (HE). These methods aim to obfuscate execution patterns and data processing sequences, complicating an attacker’s ability to correlate observable side-channel data—such as power consumption, electromagnetic emanations, or timing variations—with sensitive cryptographic operations \cite{Liptak2022}.

\begin{itemize}
    \item \textbf{Shuffling:} Shuffling involves executing operations or processing data in a randomized order \cite{Patranabis2016Shuffling}. For example, in HE systems that handle multiple ciphertexts or compute on batched data, shuffling the order of operations can disrupt predictable patterns that attackers rely on to analyze side-channel data. This technique is particularly useful in thwarting statistical side-channel attacks, where repeated patterns are exploited over multiple traces to infer secret information.  

    \item \textbf{Random Dummy Operations:}  
    Inserting random, non-functional operations (dummy computations) during cryptographic processing adds noise to power or timing traces, making it harder for attackers to distinguish real computations \cite{Liptak2022}. In HE, dummy operations must be designed to avoid disrupting the correctness of computations, as the deterministic nature of HE schemes leaves little room for errors introduced by excessive obfuscation. Practical implementations need to balance security gains with the additional computational burden. 
    
    \item \textbf{Randomized Noise Addition:}  
    Adding random noise to intermediate values during HE operations can obscure side-channel data, reducing the risk of pattern detection. However, in HE systems, this noise must be carefully managed to avoid interfering with the decryption process or exceeding the noise budget inherent to HE ciphertexts. Unlike masking, which binds randomness to cryptographic parameters, randomized noise in HE should be lightweight and consistent with the system's correctness constraints.  
    
    \item \textbf{Obfuscated Memory Access Patterns:}  
    Randomizing memory access patterns prevents attackers from exploiting cache-timing or memory-based side-channel vulnerabilities \cite{jiang2020mempoline}. In HE implementations, this could involve accessing memory blocks in a randomized order or using constant-time memory access strategies to eliminate data-dependent variations. For example, during ciphertext storage or retrieval, randomized access can ensure that memory usage does not reveal sensitive information, albeit at the cost of increased memory latency.  
\end{itemize}

While these techniques significantly enhance side-channel resilience, their implementation in HE systems introduces trade-offs between security and performance. HE operations are computationally intensive, and adding randomization mechanisms can further increase processing time and resource usage. As such, their application should be guided by careful profiling and evaluation of security versus efficiency.

\bibliographystyle{IEEEtran}
\bibliography{main}

\begin{thebibliography}{10}
\providecommand{\url}[1]{#1}
\csname url@samestyle\endcsname
\providecommand{\newblock}{\relax}
\providecommand{\bibinfo}[2]{#2}
\providecommand{\BIBentrySTDinterwordspacing}{\spaceskip=0pt\relax}
\providecommand{\BIBentryALTinterwordstretchfactor}{4}
\providecommand{\BIBentryALTinterwordspacing}{\spaceskip=\fontdimen2\font plus
\BIBentryALTinterwordstretchfactor\fontdimen3\font minus \fontdimen4\font\relax}
\providecommand{\BIBforeignlanguage}[2]{{%
\expandafter\ifx\csname l@#1\endcsname\relax
\typeout{** WARNING: IEEEtran.bst: No hyphenation pattern has been}%
\typeout{** loaded for the language `#1'. Using the pattern for}%
\typeout{** the default language instead.}%
\else
\language=\csname l@#1\endcsname
\fi
#2}}
\providecommand{\BIBdecl}{\relax}
\BIBdecl

\bibitem{paillier1999public}
P.~Paillier, ``Public-key cryptosystems based on composite degree residuosity classes,'' in \emph{International conference on the theory and applications of cryptographic techniques}.\hskip 1em plus 0.5em minus 0.4em\relax Springer, 1999, pp. 223--238.

\bibitem{elgamal1985public}
T.~ElGamal, ``A public key cryptosystem and a signature scheme based on discrete logarithms,'' \emph{IEEE transactions on information theory}, vol.~31, no.~4, pp. 469--472, 1985.

\bibitem{damgaard2010generalization}
I.~Damg{\aa}rd, M.~Jurik, and J.~B. Nielsen, ``A generalization of paillier’s public-key system with applications to electronic voting,'' \emph{International Journal of Information Security}, vol.~9, pp. 371--385, 2010.

\bibitem{okamoto1998new}
T.~Okamoto and S.~Uchiyama, ``A new public-key cryptosystem as secure as factoring,'' in \emph{Advances in Cryptology—EUROCRYPT'98: International Conference on the Theory and Application of Cryptographic Techniques Espoo, Finland, May 31--June 4, 1998 Proceedings 17}.\hskip 1em plus 0.5em minus 0.4em\relax Springer, 1998, pp. 308--318.

\bibitem{cohen1985robust}
J.~D. Cohen and M.~J. Fischer, \emph{A robust and verifiable cryptographically secure election scheme}.\hskip 1em plus 0.5em minus 0.4em\relax Yale University. Department of Computer Science, 1985.

\bibitem{naccache1997new}
D.~Naccache and J.~Stern, ``A new public-key cryptosystem,'' in \emph{Advances in Cryptology—EUROCRYPT’97: International Conference on the Theory and Application of Cryptographic Techniques Konstanz, Germany, May 11--15, 1997 Proceedings 16}.\hskip 1em plus 0.5em minus 0.4em\relax Springer, 1997, pp. 27--36.

\bibitem{goldwasser2019probabilistic}
S.~Goldwasser and S.~Micali, ``Probabilistic encryption \& how to play mental poker keeping secret all partial information,'' in \emph{Providing sound foundations for cryptography: on the work of Shafi Goldwasser and Silvio Micali}, 2019, pp. 173--201.

\bibitem{sealmanual}
M.~S. Team, ``Microsoft seal (simple encrypted arithmetic library),'' 2022, available at \url{https://www.microsoft.com/en-us/research/project/microsoft-seal/}.

\bibitem{openfhe_github}
O.~D. Team, ``Openfhe: Open-source fully homomorphic encryption library,'' GitHub Repository, 2023, \url{https://github.com/openfheorg/openfhe-development}.

\bibitem{Mangard2010}
S.~Mangard, E.~Oswald, and T.~Popp, \emph{Power Analysis Attacks: Revealing the Secrets of Smart Cards}, 1st~ed.\hskip 1em plus 0.5em minus 0.4em\relax Springer Publishing Company, Incorporated, 2010.

\bibitem{Picek2023}
\BIBentryALTinterwordspacing
S.~Picek, G.~Perin, L.~Mariot, L.~Wu, and L.~Batina, ``Sok: Deep learning-based physical side-channel analysis,'' \emph{ACM Computing Surveys (ACM Comput. Surv.)}, vol.~55, no.~11, pp. Article 227, 35 pages, November 2023. [Online]. Available: \url{https://doi.org/10.1145/3569577}
\BIBentrySTDinterwordspacing

\bibitem{Devi2021}
\BIBentryALTinterwordspacing
M.~Devi and A.~Majumder, ``Side-channel attack in internet of things: A survey,'' in \emph{Applications of Internet of Things}, ser. Lecture Notes in Networks and Systems, J.~Mandal, S.~Mukhopadhyay, and A.~Roy, Eds.\hskip 1em plus 0.5em minus 0.4em\relax Springer, Singapore, 2021, vol. 137, pp. 257--270. [Online]. Available: \url{https://doi.org/10.1007/978-981-15-6198-6_20}
\BIBentrySTDinterwordspacing

\bibitem{strobel2009side}
D.~Strobel, I.~C. Paar, and M.~Kasper, ``Side channel analysis attacks on stream ciphers,'' \emph{Masterarbeit Ruhr-Universit{\"a}t Bochum, Lehrstuhl Embedded Security}, 2009.

\bibitem{power_image}
\BIBentryALTinterwordspacing
{Audriusa (Wikimedia Commons)}, ``Oscilloscope reading showing power consumption variations,'' 2024, licensed under the GNU Free Documentation License (GFDL). [Online]. Available: \url{https://en.wikipedia.org/wiki/File:Image.png}
\BIBentrySTDinterwordspacing

\bibitem{Aydin2022}
F.~Aydin, E.~Karabulut, S.~Potluri, E.~Alkim, and A.~Aysu, ``Reveal: Single-trace side-channel leakage of the seal homomorphic encryption library,'' in \emph{2022 Design, Automation \& Test in Europe Conference \& Exhibition (DATE)}, Antwerp, Belgium, 2022, pp. 1527--1532.

\bibitem{Onishi2024}
\BIBentryALTinterwordspacing
R.~Onishi, T.~Suzuki, S.~Sakai, and H.~Yamana, ``Security and performance-aware cloud computing with homomorphic encryption and trusted execution environment,'' in \emph{Proceedings of the 12th Workshop on Encrypted Computing \& Applied Homomorphic Cryptography (WAHC '24)}.\hskip 1em plus 0.5em minus 0.4em\relax New York, NY, USA: Association for Computing Machinery, 2024, pp. 36--42. [Online]. Available: \url{https://doi.org/10.1145/3689945.3694805}
\BIBentrySTDinterwordspacing

\bibitem{rivest1978data}
R.~L. Rivest, L.~Adleman, M.~L. Dertouzos \emph{et~al.}, ``On data banks and privacy homomorphisms,'' \emph{Foundations of secure computation}, vol.~4, no.~11, pp. 169--180, 1978.

\bibitem{asecuritysite_17070}
\BIBentryALTinterwordspacing
W.~J. Buchanan, ``Openfhe,'' \url{https://github.com/openfheorg/openfhe-development}, OpenFHE, 2024, accessed: Feb 20, 2025. [Online]. Available: \url{https://github.com/openfheorg/openfhe-development}
\BIBentrySTDinterwordspacing

\bibitem{homenc}
C.~Gentry, ``A fully homomorphic encryption scheme,'' 2009, \url{crypto.stanford.edu/craig}.

\bibitem{van2010fully}
M.~Van~Dijk, C.~Gentry, S.~Halevi, and V.~Vaikuntanathan, ``Fully homomorphic encryption over the integers,'' in \emph{Advances in Cryptology--EUROCRYPT 2010: 29th Annual International Conference on the Theory and Applications of Cryptographic Techniques, French Riviera, May 30--June 3, 2010. Proceedings 29}.\hskip 1em plus 0.5em minus 0.4em\relax Springer, 2010, pp. 24--43.

\bibitem{brakerski2014efficient}
Z.~Brakerski and V.~Vaikuntanathan, ``Efficient fully homomorphic encryption from (standard) lwe,'' \emph{SIAM Journal on computing}, vol.~43, no.~2, pp. 831--871, 2014.

\bibitem{cheon2017homomorphic}
J.~H. Cheon, A.~Kim, M.~Kim, and Y.~Song, ``Homomorphic encryption for arithmetic of approximate numbers,'' in \emph{Advances in Cryptology--ASIACRYPT 2017: 23rd International Conference on the Theory and Applications of Cryptology and Information Security, Hong Kong, China, December 3-7, 2017, Proceedings, Part I 23}.\hskip 1em plus 0.5em minus 0.4em\relax Springer, 2017, pp. 409--437.

\bibitem{asecuritysite_85691}
\BIBentryALTinterwordspacing
W.~J. Buchanan, ``Homomorphic encryption (seal),'' \url{https://asecuritysite.com/seal}, Asecuritysite.com, 2024, accessed: September 04, 2024. [Online]. Available: \url{https://asecuritysite.com/seal}
\BIBentrySTDinterwordspacing

\bibitem{asecuritysite_40933}
\BIBentryALTinterwordspacing
------, ``Homomorphic encryption with bfv using node.js,'' \url{https://asecuritysite.com/seal/js_homomorphic}, Asecuritysite.com, 2025, accessed: February 28, 2025. [Online]. Available: \url{https://asecuritysite.com/seal/js_homomorphic}
\BIBentrySTDinterwordspacing

\bibitem{wood2020homomorphic}
A.~Wood, K.~Najarian, and D.~Kahrobaei, ``Homomorphic encryption for machine learning in medicine and bioinformatics,'' \emph{ACM Computing Surveys (CSUR)}, vol.~53, no.~4, pp. 1--35, 2020.

\bibitem{ducas2015fhew}
L.~Ducas and D.~Micciancio, ``Fhew: bootstrapping homomorphic encryption in less than a second,'' in \emph{Annual international conference on the theory and applications of cryptographic techniques}.\hskip 1em plus 0.5em minus 0.4em\relax Springer, 2015, pp. 617--640.

\bibitem{al2023demystifying}
A.~Al~Badawi and Y.~Polyakov, ``Demystifying bootstrapping in fully homomorphic encryption,'' \emph{Cryptology ePrint Archive}, 2023.

\bibitem{asecuritysite_65179}
\BIBentryALTinterwordspacing
W.~J. Buchanan, ``Chebyshev approximations using openfhe and c++ (logarithm methods),'' \url{https://asecuritysite.com/openfhe/openfhe_18cpp}, Asecuritysite.com, 2024, accessed: September 04, 2024. [Online]. Available: \url{https://asecuritysite.com/openfhe/openfhe_18cpp}
\BIBentrySTDinterwordspacing

\bibitem{Spreitzer2018}
R.~Spreitzer, V.~Moonsamy, T.~Korak, and S.~Mangard, ``Systematic classification of side-channel attacks: A case study for mobile devices,'' \emph{IEEE Communications Surveys \& Tutorials}, vol.~20, no.~1, pp. 465--488, Firstquarter 2018.

\bibitem{Standaert2010}
\BIBentryALTinterwordspacing
F.-X. Standaert, \emph{Introduction to Side-Channel Attacks}.\hskip 1em plus 0.5em minus 0.4em\relax Boston, MA: Springer US, 2010, pp. 27--42. [Online]. Available: \url{https://doi.org/10.1007/978-0-387-71829-3_2}
\BIBentrySTDinterwordspacing

\bibitem{Messerges2000}
\BIBentryALTinterwordspacing
T.~Messerges, ``Using second-order power analysis to attack dpa resistant software,'' in \emph{Cryptographic Hardware and Embedded Systems -- CHES 2000}, ser. Lecture Notes in Computer Science.\hskip 1em plus 0.5em minus 0.4em\relax Springer, Berlin, Heidelberg, 2000, vol. 1965, pp. 238--251. [Online]. Available: \url{https://doi.org/10.1007/3-540-44499-8_19}
\BIBentrySTDinterwordspacing

\bibitem{Aydin2024}
\BIBentryALTinterwordspacing
F.~Aydin and A.~Aysu, ``Leaking secrets in homomorphic encryption with side-channel attacks,'' \emph{Journal of Cryptographic Engineering (J Cryptogr Eng)}, vol.~14, pp. 241--251, 2024. [Online]. Available: \url{https://doi.org/10.1007/s13389-023-00340-2}
\BIBentrySTDinterwordspacing

\bibitem{Ravi2022}
P.~Ravi, S.~Bhasin, S.~S. Roy, and A.~Chattopadhyay, ``On exploiting message leakage in (few) nist pqc candidates for practical message recovery attacks,'' \emph{IEEE Transactions on Information Forensics and Security}, vol.~17, pp. 684--699, 2022.

\bibitem{Cheng2022}
\BIBentryALTinterwordspacing
W.~Cheng, J.-L. Danger, S.~Guilley, F.~Huang, A.~B. Korchi \emph{et~al.}, ``Cache-timing attack on the seal homomorphic encryption library,'' in \emph{11th International Workshop on Security Proofs for Embedded Systems (PROOFS 2022)}, Leuven, Belgium, Sep 2022. [Online]. Available: \url{https://hal.archives-ouvertes.fr/hal-03780506}
\BIBentrySTDinterwordspacing

\bibitem{Primas2017}
\BIBentryALTinterwordspacing
R.~Primas, P.~Pessl, and S.~Mangard, ``Single-trace side-channel attacks on masked lattice-based encryption,'' in \emph{Cryptographic Hardware and Embedded Systems -- CHES 2017}, ser. Lecture Notes in Computer Science, W.~Fischer and N.~Homma, Eds., vol. 10529.\hskip 1em plus 0.5em minus 0.4em\relax Springer, Cham, 2017, pp. 537--557. [Online]. Available: \url{https://doi.org/10.1007/978-3-319-66787-4_25}
\BIBentrySTDinterwordspacing

\bibitem{Chaturvedi2022}
\BIBentryALTinterwordspacing
B.~Chaturvedi, A.~Chakraborty, A.~Chatterjee, and D.~Mukhopadhyay, ``A practical full key recovery attack on tfhe and fhew by inducing decryption errors,'' \emph{Cryptology ePrint Archive}, 2022. [Online]. Available: \url{https://eprint.iacr.org/2022/1563}
\BIBentrySTDinterwordspacing

\bibitem{Lyu2018}
\BIBentryALTinterwordspacing
Y.~Lyu and P.~Mishra, ``A survey of side-channel attacks on caches and countermeasures,'' \emph{Journal of Hardware and Systems Security (J Hardw Syst Secur)}, vol.~2, pp. 33--50, 2018. [Online]. Available: \url{https://doi.org/10.1007/s41635-017-0025-y}
\BIBentrySTDinterwordspacing

\bibitem{Cock2014}
\BIBentryALTinterwordspacing
D.~Cock, Q.~Ge, T.~Murray, and G.~Heiser, ``The last mile: An empirical study of timing channels on sel4,'' in \emph{Proceedings of the 2014 ACM SIGSAC Conference on Computer and Communications Security (CCS '14)}.\hskip 1em plus 0.5em minus 0.4em\relax New York, NY, USA: Association for Computing Machinery, 2014, pp. 570--581. [Online]. Available: \url{https://doi.org/10.1145/2660267.2660294}
\BIBentrySTDinterwordspacing

\bibitem{Hou2024}
\BIBentryALTinterwordspacing
X.~Hou and J.~Breier, ``Side-channel analysis attacks and countermeasures,'' in \emph{Cryptography and Embedded Systems Security}.\hskip 1em plus 0.5em minus 0.4em\relax Springer, Cham, 2024. [Online]. Available: \url{https://doi.org/10.1007/978-3-031-62205-2_4}
\BIBentrySTDinterwordspacing

\bibitem{Patranabis2016Shuffling}
S.~Patranabis, D.~B. Roy, P.~K. Vadnala, D.~Mukhopadhyay, and S.~Ghosh, ``Shuffling across rounds: A lightweight strategy to counter side-channel attacks,'' in \emph{2016 IEEE 34th International Conference on Computer Design (ICCD)}, 2016, pp. 440--443.

\bibitem{Liptak2022}
\BIBentryALTinterwordspacing
C.~Liptak, S.~Mal-Sarkar, and S.~A.~P. Kumar, ``Power analysis side channel attacks and countermeasures for the internet of things,'' in \emph{2022 IEEE Physical Assurance and Inspection of Electronics (PAINE)}.\hskip 1em plus 0.5em minus 0.4em\relax IEEE, 2022, pp. 1--7. [Online]. Available: \url{https://doi.org/10.1109/PAINE56030.2022.10014854}
\BIBentrySTDinterwordspacing

\bibitem{jiang2020mempoline}
\BIBentryALTinterwordspacing
Z.~H. Jiang, Y.~Fei, A.~A. Ding, and T.~Wahl, ``Mempoline: Mitigating memory-based side-channel attacks through memory access obfuscation,'' \emph{Cryptology ePrint Archive}, 2020. [Online]. Available: \url{https://eprint.iacr.org/2020/760.pdf}
\BIBentrySTDinterwordspacing

\end{thebibliography}

\end{document}